\documentclass[aip,apl,amsmath,reprint,amssymb,floatfix]{revtex4-1} 
\usepackage{graphicx}
\usepackage{dcolumn}

\begin{document}

\title{Radio-frequency reflectometry on an undoped AlGaAs/GaAs single electron transistor}

\author{S. J. MacLeod}
 \affiliation{School of Physics, University of New South Wales, Sydney, New South Wales 2052, Australia}
\author{A. M. See}
 \affiliation{School of Physics, University of New South Wales, Sydney, New South Wales 2052, Australia}
\author{Z. K. Keane}
 \altaffiliation[Now at ]{Northrop Grumman Electronic Systems, Linthicum, Maryland 21090, USA}
 \affiliation{School of Physics, University of New South Wales, Sydney, New South Wales 2052, Australia}
\author{P. Scriven}
 \affiliation{School of Physics, University of New South Wales, Sydney, New South Wales 2052, Australia}
\author{A. P. Micolich}
 \affiliation{School of Physics, University of New South Wales, Sydney, New South Wales 2052, Australia}
\author{M. Aagesen}
\affiliation{Nanoscience center, University of Copenhagen, Universitetsparken 5, DK-2100 Copenhagen, Denmark.}
\author{P. E. Lindelof}
\affiliation{Nanoscience center, University of Copenhagen, Universitetsparken 5, DK-2100 Copenhagen, Denmark.}
\author{A. R. Hamilton}
 \email{Alex.Hamilton@unsw.edu.au}
 \affiliation{School of Physics, University of New South Wales, Sydney, New South Wales 2052, Australia}

\date{\today}
\begin{abstract}
Radio frequency reflectometry is demonstrated in a sub-micron undoped AlGaAs/GaAs device. Undoped single electron transistors (SETs) are attractive candidates to study single electron phenomena due to their charge stability and robust electronic properties after thermal cycling. However these devices require a large top-gate which is unsuitable for the fast and sensitive radio frequency reflectometry technique. Here we demonstrate rf reflectometry is possible in an undoped SET.
\end{abstract}


\maketitle
Typically AlGaAs/GaAs systems are realised by modulation doping: a layer of dopants donate excess charge carriers to form a two-dimensional electron gas (2DEG) at the AlGaAs/GaAs interface. Small metallic surface gates are used to define sub-micron features: applying a voltage depletes the 2DEG in the regions under the gates. Modulation doped devices have been integrated into radio-frequency (rf) resonant circuits~\cite{Schoelkopf1998}, because they have small parasitic capacitances~\cite{Aassime2001,Xue2009,Lu2003} which are used to form an $LC$ resonant circuit when in series with an inductor~\cite{Schoelkopf1998}. Radio frequency techniques are well suited to exploring phenomena over short time scales, and in modulation doped devices, are central to experiments on qubit entanglement~\cite{Shulman2012}, noise~\cite{Fujisawa2000} and spin manipulation~\cite{Laird2010}. 

\begin{figure}[ht]
        \includegraphics[width=0.36\textwidth]{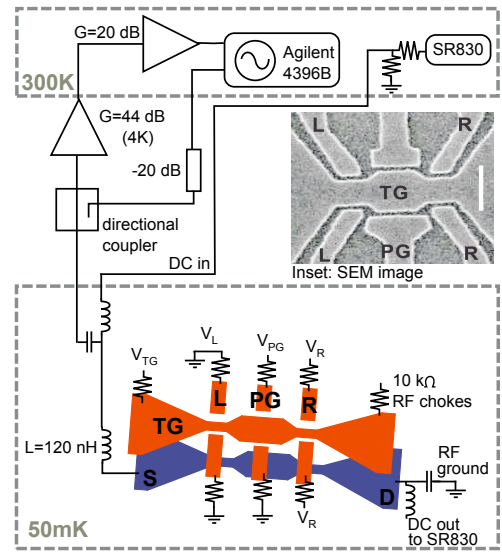}
        \caption{(color online) The undoped SET is embedded in a measurement set-up capable of simultaneous rf and dc measurements. The 2DEG is induced by applying a voltage to the top-gate (TG) with current passed between the source (S) and drain (D). The SET is controlled by applying voltage to the left- (L), right- (R), and plunger-gate (PG) gates via 10\,k$\Omega$ rf-chokes (L was kept grounded due to strong coupling with the PG). Inset, a SEM image of the SET, the darker regions are where the top-gate has been etched away. The scale bar is $500\,\mathrm{nm}$.}
        \label{fig:ExpSetup}
    \end{figure}

As an alternative, undoped AlGaAs/GaAs devices can be thought of as the inverse of modulation doped devices. A voltage is applied to an overall top-gate inducing a 2DEG at the AlGaAs/GaAs interface. If the gate is an epitaxial, degenerately doped GaAs layer~\cite{See2010} then sub-micron structures, such as single electron transistors (SETs), are realised in the 2DEG by etching away small regions of the top-gate~\cite{Kane1998,Klochan2006,See2010} (inset to Figure~\ref{fig:ExpSetup}). Sub-micron, undoped devices have shown excellent charge stability~\cite{Klochan2006}, as well as thermal robustness~\cite{See2010}. In comparison, modulation doped structures can suffer from significant charge noise due to surface gate leakage~\cite{Buizert2008,Pioro-Ladriere2005,Cobden1992} and fluctuators in the doping layer~\cite{Fujisawa2000,Kurdak1997,Timp1990}. Charge sensing using rf techniques has yet to be used with sub-micron, undoped devices, so probing processes over shorter time scales or fast charge sensing applications, remain unstudied in these systems.
%
%

In this paper we address the question: is it possible to use rf reflectometry in a sub-micron, undoped AlGaAs/GaAs SET? The SET is formed in a 2D conducting channel between the source and drain. The 2D leads, as well as the SET are covered by the top-gate. Of central concern is the effect of the large top-gate capacitance on the reflected rf signal. The high-frequency signal could be shunted by the top-gate before reaching the SET, and the large top-gate capacitance could make impedance matching difficult at high frequencies~\cite{Taskinen2008}. Reducing the top-gate capacitance by shrinking the area may not improve the impedance matching of the $LC$ circuit. This is because the gated 2DEG acts as a lossy transmission line so that it is the capacitance per unit area, not the total capacitance, which determines the high frequency behaviour of the device~\cite{Taskinen2008}. Furthermore, as the SET is only a small structure embedded in a 2D conducting channel, it is unclear how sensitive the rf reflectance ($S_{11}$) will be to the SET resistance.

The device used here\cite{See2010} consists of a 35\,nm-thick GaAs n$+$ cap (the top-gate)~ covering the Hall bar and has an area of $\sim 2.78\times 10^{-7}\,\mathrm{m^{2}}$. Applying a positive voltage to the top-gate controls the electron density, $n$, of the 2DEG which resides 185\,nm below the cap, with a geometric capacitance $\sim 169.2$\,pF to the top-gate. The SET is defined by etching the $n+$ cap into seven separate gates, labelled left- (L), right- (R), plunger- (PG) and top-gate (TG) as shown in Figure~\ref{fig:ExpSetup}, the capacitances are obtained from previous measurements~\cite{See2010} and are 17.0, 14.5, 20 and 107\,aF respectively~\cite{See2010}. To define the SET a negative voltage is applied to the left-,right- and plunger-gates ($V_{L}$, $V_{R}$ and $V_{PG}$ respectively), thus controlling the tunnel barrier heights and the electron number in the SET. The charging energy of the SET is $E_{C}=1\,\mathrm{meV}$~\cite{See2010}.

Low temperature measurements were made in a dilution refrigerator with a 20\,mK base temperature. The rf impedance matching network is formed by soldering a $120\,\mathrm{nH}$ inductor onto a printed circuit board (PCB) onto which the sample, bonded in a standard LCC20 chip package, is mounted. The low-frequency lines have $10\,\mathrm{k\Omega}$ rf-chokes soldered in place. Electrical connection between the chip package and PCB is via 20$\,\mathrm{\mu m}$ gold-bonding wires. Figure~\ref{fig:ExpSetup} shows the low-temperature set up for recovering our reflected signal: once the carrier wave is reflected it is separated from the incoming wave via a directional coupler and amplified with a CITLF1 low-noise broadband amplifier (44\,dB gain at 20\,K) operating at 4\,K. This is followed by a 300\,K amplification stage (mini-circuits ZFL-1000LN) with 20\,dB of gain. The reflected power is measured with an Agilent rf spectrum analyzer or a square-law diode detector (TruPwr Detector, model AD8362), with data averaged over four traces.

\begin{figure}
        \includegraphics[width=0.38\textwidth]{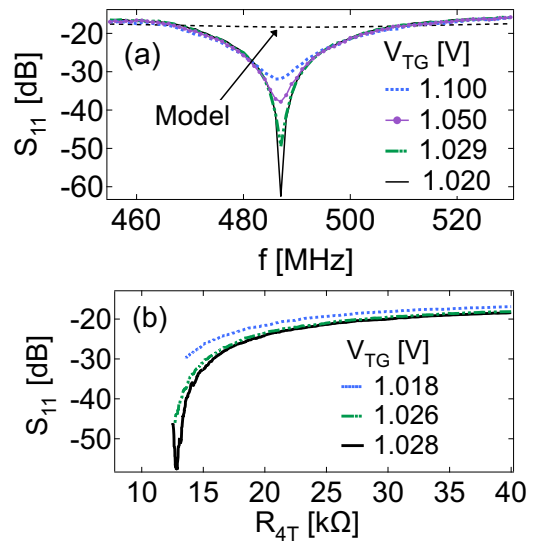}
        \caption{(color online) 4\,K characterisation of the SET. (a) The measured dependence of rf reflectance $S_{11}$
 on frequency ($f$) is shown for different top-gate voltages $V_{TG}=\{1.020,\,1.029,\,1.050,\,1.100\}\,\mathrm{V}$ corresponding to electron densities ($n$) of $\{2.39,\,2.42,\,2.49,\,2.66\}\times 10^{11}\,\mathrm{cm^{-2}}$ respectively. $S_{11}$ starts at -17 dB due to losses in the lines from the top of the rf
dip-station to the sample stage at room temperature. The dashed black line is the result of a lumped-element circuit model~\cite{InfoE} for $V_{TG}=1.020$\,V. (b) The rf reflectometry circuit is at resonance (487\,MHz). We plot $S_{11}$ as a function of the four-terminal resistance of the dot, where $R_{4T}$ is varied by changing the plunger gate voltage $V_{PG}$. Measurements were performed for
$V_{TG}=\{1.018,\,1.026,\,1.028\}\,\mathrm{V}$, where $n=\{2.38,\,2.41,\,2.42\} \times 10^{11}\,\mathrm{cm^{2}}$.~\cite{See2010}}
        \label{fig:TransFnc}
    \end{figure}

The device is first characterised at 4\,K in a dedicated rf dip-station with similar wiring to the dilution fridge. To locate the resonant frequency and check $S_{11}$ is sensitive to the sample resistance, the frequency is swept while $S_{11}$ is recorded for different values of $V_{TG}$ when the dot is unbiased. Sweeping the frequency yields a sharp resonance at 487\,MHz, with a 100\,nH, inductor as shown in Figure~\ref{fig:TransFnc}(a). There is a change in $S_{11}$ of up to 40\,dB depending on $V_{TG}$. The $S_{11}$ versus frequency curves are determined by the matching of the sample resistance (set here by $V_{TG}$) to the $LC$ circuit. Thus the depth of the $S_{11}$ minima depends on $V_{TG}$ because the resistance per square of the Hall bar and the SET are controlled by $V_{TG}$. We have also used a simple lumped-element analysis to model the circuit, as shown by the upper trace in Fig.~\ref{fig:TransFnc}(a). Although this simple model works well for 2D systems~\cite{Taskinen2008}, it cannot describe the experimental data when the dot is included. More sophisticated modelling should be able to replicate the experimental data, but requires many poorly known parameters, thus limiting the ability to predict the circuit response.
%
%

To isolate the effect of the SET we set the signal frequency on resonance (487\,MHz) and record $S_{11}$ versus the four-terminal resistance ($R_{4T}$) across the SET; $R_{4T}$ is changed by sweeping $V_{PG}$ while the top-, left- and right-gate voltages are fixed~\cite{InfoA}. The transfer function is shown in Figure~\ref{fig:TransFnc}(b), where $S_{11}$ versus $R_{4T}$ is plotted for different values of $V_{TG}$. For $V_{TG}=1.028$\,V the circuit is matched at $R_{4T}=30\,\mathrm{k\Omega}$, and $S_{11}$ shows a change of up to 40\,dB as $R_{4T}$ increases. For lower $V_{TG}$ the matching point moves to lower $R_{4T}$.
%
%
    \begin{figure}
        \includegraphics[width=0.45\textwidth]{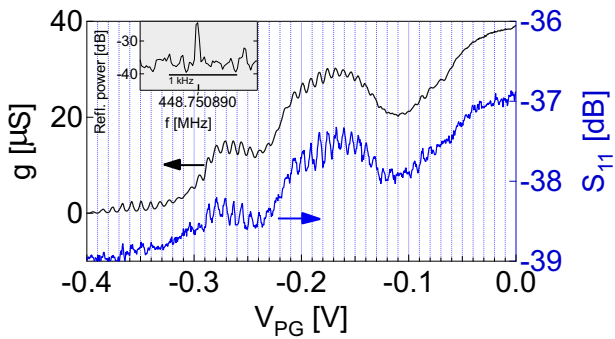}
        \caption{(color online) Millikelvin measurements of the Coulomb blockade (CB) oscillations (left axis) in the two-terminal conductance $g$ (series resistance subtracted~\cite{InfoD}). The rf-reflectance $S_{11}$ (right axis) traces out the CB oscillations when the transfer function is sensitive to the dot. We apply $-50\,\mathrm{dBm}$ at 447\,MHz to the source. Losses in the lines from the top of the dilution fridge to the sample stage at room temperature are -37\,dB marking the upper limit of $S_{11}$. The inset shows the spectrum of the reflected power for a small ($0.09 e_{rms}$.) 890\,Hz sine wave on the top plunger-gate.}
        \label{fig:CBwithRF}
    \end{figure}

The response of $S_{11}$ to the SET is best demonstrated by mapping out the conductance oscillations when the SET~\cite{InfoB} is in the Coulomb blockade (CB) regime at low temperature. To improve matching the inductor is changed to 120\,nH and the circuit resonance frequency becomes 448.75\,MHz. Figure~\ref{fig:CBwithRF} shows the CB oscillations in $g$ measured using the low-frequency, two-terminal lockin technique. The CB oscillations are reproduced in $S_{11}$ when $1/g\sim 33\,\mathrm{k\Omega}-100\,\mathrm{k\Omega}$. Here the change in the amplitude of $S_{11}$ detecting the CB oscillations is small because the sample resistance is larger than $20\,\mathrm{k\Omega}$. From Figure~\ref{fig:TransFnc}(b) we see if the sample resistance varies between $20\,\mathrm{k\Omega}$ to $>40\,\mathrm{k\Omega}$ then $S_{11}$ changes by $-5$\,dB or less. Thus for a sample resistance between $33\,\mathrm{k\Omega}-100\,\mathrm{k\Omega}$ the change in amplitude of $S_{11}$ is small. The best charge sensitivity is obtained at the maximal slope $\partial S_{11}/\partial V_{PG}$, thus we configure the dot to sit on the side of a CB peak and apply a small signal excitation ($\sim 0.09\,\mathrm{e}$) to the top plunger gate at 890\,Hz. We obtain a charge sensitivity of $\delta q = 2.6 \times 10^{-3}\,\mathrm{e/\sqrt{Hz}}$ with a 3\,Hz resolution bandwidth (see inset to Fig.3). We note this charge sensitivity is likely an underestimate, since the SET is still affected by $1/f$ charge noise, as we were limited by gate modulation frequencies below the $1/f$ knee due to low-frequency gate lines. Furthermore the rf matching is far from optimized, so there is much room for improvement which is important for future applications of single shot read-out of undoped quantum dots for spin qubits~\cite{Barthel2010}. One way to improve the rf matching is to iteratively try different inductors, requiring a thermal cycle with each attempt. Another method is to use a variable capacitor to tune the resonance frequency \textit{in situ}~\cite{Muller2010}. 
%
%
\begin{figure}
    \includegraphics[width=0.45\textwidth]{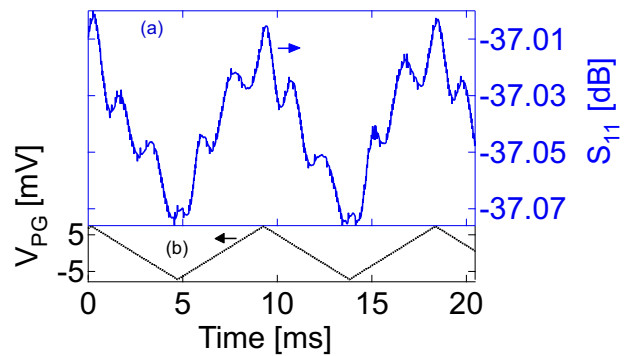}
    \caption{(color online) Rapid gated measurements of Coulomb blockade (CB) oscillations. The SET is biased to sit on the side of a CB peak and a 15\,mV triangle-wave is applied at 110\,Hz to the top plunger-gate. (a) $S_{11}$ monitors the changing occupation number of the SET as indicated by the smaller peaks, and (b) shows the triangle wave applied to the top plunger-gate.}
\label{fig:FastCB}
\end{figure}

We now demonstrate rf reflectometry by using $S_{11}$ to track changes in the occupation number of the SET. The SET is configured~\cite{InfoC} in the CB regime while a 15\,mV, 110\,Hz triangle wave (Figure~\ref{fig:FastCB}(b)) is applied to the top half of the plunger-gate and $S_{11}$ is recorded versus time (Figure~\ref{fig:FastCB}(a)). The smaller peaks in $S_{11}$, spaced 1.5\,ms apart, show the occupation number of the SET changing as $V_{PG}$ is swept. This corresponds to the SET passing through $\sim$1000 CB peaks a second. The frequency applied to the gate is limited by low-pass filters on the gate lines $-$ without these the gate could be modulated at much higher frequencies. 

In conclusion we have shown rf reflectometry is possible on an undoped, large-area gated, sub-micron SET, despite the predictions of a simple lumped-element circuit model. We observed clear Coulomb blockade in $S_{11}$. Future work will include optimising the rf matching and improving the charge sensitivity of the undoped quantum dot.

This work was funded by the Australian Research Council (ARC) DP and FT schemes. A.R.H. acknowledges an ARC Outstanding Researcher Award.
%
%
\bibliographystyle{apsrev4-1}

\begin{thebibliography}{23}
\bibitem {InfoD} {T}he series resistance is $\sim 23.4\,\mathrm{k\Omega}$ due to: rf chokes ($10\,\mathrm{k\Omega}\times 2$); ohmic contacts ($1\,\mathrm{k\Omega}\times 2$); line resistance ($200\,\mathrm{\Omega}\times 2$) and the lock-in input mpedance ($1\,\mathrm{k\Omega}$).
\bibitem{InfoB} {T}he {SET} was biased with $V_{L}=0\,\mathrm{V}$, $V_{R}=-0.11\,\mathrm{V}$, and sweeping  $V_{PG}$($-0.5\,\mathrm{V}<V_{PG}<0\,\mathrm{V}$. {T}he best results for this configuration were obtained when $V_{TG}=1\,\mathrm{V}$.
\end{thebibliography}
%
%

%

\end{document}